\documentclass[aps,prl,twocolumn,preprintnumbers,showpacs,amsmath,amssymb,floatfix]{revtex4}

\usepackage{graphicx}
\usepackage{bm} 

\begin{document}
\preprint{{\em Submitted to Phys. Rev. Lett. (2005)}}

\title{Counterions at Charged Cylinders: Criticality and universality beyond mean-field}

\author{Ali Naji}
\author{Roland R. Netz}

\affiliation{Physics Department, Technical University of Munich, 85748 Garching, Germany.}
                 
\date{April 2005}

\begin{abstract}
The counterion-condensation transition at charged cylinders is studied using
Monte-Carlo simulation methods. Employing logarithmically rescaled 
radial coordinates, large system sizes are tractable and the 
critical behavior is determined by a combined finite-size and finite-ion-number 
analysis. Critical counterion localization exponents are introduced
and found to be in accord with  mean-field theory both in 2 and 3 dimensions.
In 3D the heat capacity shows a universal jump at the transition, while in 
2D, it consists of discrete peaks where single counterions successively condense.
\end{abstract}

\pacs{
64.60.Fr,  
82.35.Rs, 
87.15.-v,  
61.20.Ja  
         }

\maketitle


Many biopolymers, such as DNA, actin, tubulin, fd-viruses,
are charged and stiff. On length scales smaller than the persistence length,
they can be represented by straight, charged cylinders and oppositely 
charged ions (counterions) are attracted via an electrostatic potential that grows 
logarithmically with radial distance. As the ion confinement entropy 
also exhibits a logarithmic dependence, it was suggested early by Onsager that 
a counterion delocalization transition occurs at a critical cylinder charge 
(or equivalently at a critical temperature)  \cite{Manning}.
This argument is strictly valid only for a single ion since it neglects cooperativity due
to inter-ionic repulsions.  Nevertheless, it was corroborated by 
mean-field (MF) studies \cite{Manning,Alfrey,Zimm}, which demonstrate that 
below a critical temperature, a fraction of counterions 
stays bound or condensed in the vicinity of the central cylinder even in the limit of 
infinite system size;  while above the critical temperature, all counterions
de-condense to infinity. 
This {\em counterion-condensation transition} (CCT) 
dramatically affects a whole number of static and dynamic quantities 
for charged polymers \cite{Manning}.  It has been observed with different  polymers
by varying the medium dielectric constant \cite{Klein,Hoagland03} or
the polymer charge density \cite{Hoagland03,Ander_Penafiel};
the counterion distribution around DNA strands
has been directly measured
recently using anomalous scattering techniques \cite{Das}.
Since its discovery, the CCT has been at the focus of
numerical \cite{Deserno,Liao} and analytical  \cite{theory} studies.
Under particular dispute  has been the connection between CCT and the celebrated 
Kosterlitz-Thouless transition of logarithmically interacting particles in 
2 dimensions \cite{KT}. Also, 
the precise location of the CCT critical point remains subject of ongoing experimental 
investigations \cite{Hoagland03,Ander_Penafiel}.

As is well known from bulk critical phenomena, fluctuations and correlations
typically make non-universal and universal quantities
deviate from mean-field theory (MFT) below the upper critical
dimension \cite{critical}. Surprisingly, the MFT prediction for the CCT
critical temperature has not been questioned in literature
and apparently assumed to be exact. 
Likewise, the existence of scaling relations and critical exponents
associated with the CCT has not been considered, neither on the MFT level
and (consequently) also not in the presence  of correlations.

In this paper we pose the questions:  i) what is the critical temperature of the CCT,
and ii) what are the associated relevant critical exponents?
We employ  Monte-Carlo simulations,
which are performed in rescaled logarithmic coordinates
in order to handle very large systems (where the criticality actually occurs)
with tractable equilibration times.  A combined finite-size and
finite-ion-number analysis yields the desired critical temperature and exponents.
To enhance  the effects of fluctuations, we also study
a (within MFT equivalent) 2D system of logarithmically interacting charges, as
applicable to an experimental system of oriented
cationic and anionic polymers (e.g. DNA with polylysine \cite{Jason}). 
Surprisingly, MFT is demonstrated
to be accurate both in 3D (charged cylinder with point-like counterions)
and in 2D (charged cylinder with cylindrical counterions).
The critical exponents associated with the inverse
counterion localization length
(which plays the order parameter of the CCT) and the 
universal behavior of the heat capacity are determined;
both quantities are experimentally accessible.

\begin{figure*}[t]
\includegraphics[width=14.cm]{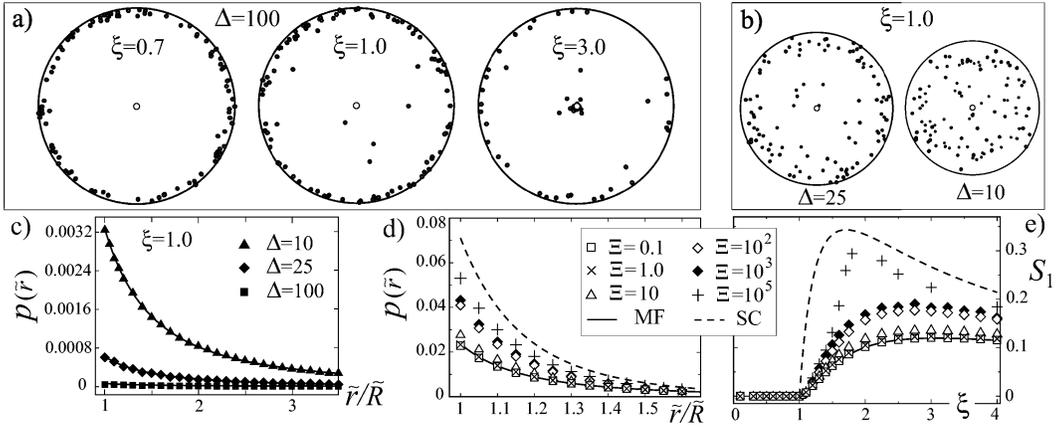}
\caption{Snapshot-topviews  for a) logarithmic box radius 
$\Delta= \ln (D/R) =100$ and Manning parameters 
$\xi=0.7, 1.0$ and 3.0, and b)  $\xi=1.0$ and  $\Delta=10$ and 25
shown in logarithmic radial units $y=\ln( r/R)$
(number of particles $N=100$ and the coupling parameter $\Xi=0.1$).
c) Radial counterion  distribution function, $ p(\tilde r)$, 
for  $\xi=1.0$,  $\Xi=0.1$, $N=100$ and various system sizes.
d) Counterion distribution for $\xi=3.0$ and 
various  $\Xi$ shows a crossover between
 MFT (solid curve) and SC (dashed curve) predictions,
Eqs. (\ref{eq:rho_MF}) and (\ref{eq:rho_SC}).
e) Order parameter $S_1=\langle 1/{\tilde r}\rangle$ as a function of 
Manning parameter, $\xi$, for different coupling $\Xi$ compared with MFT (solid curve) 
and SC (dashed curve) results,  Eqs. (\ref{eq:S_n_MF}) and (\ref{eq:Sn_SC_full}).
In d), e) $\Delta=300$ and typically $N=200$.}
\label{fig:fig1}
\vspace{-0.4cm}
\end{figure*}

In the 3D simulations, we consider a 
central  cylinder of radius $R$ 
and uniform surface charge density $\sigma_{\text{s}}$ 
(linear charge density $\tau=2\pi R\sigma_{\text{s}}$) 
with  $N$ neutralizing point-like counterions of valency $q$
confined laterally in an outer cylindrical box of radius $D$ 
(see Fig.~\ref{fig:fig1} for snapshots projected along the $z$-axis).
Periodic boundary conditions in $z$-direction are handled 
using summation methods for long-range interactions \cite{Sperb}.
Rescaling all spatial coordinates by the Gouy-Chapman length,
 $\mu=1/(2\pi q \ell_{\text{B}} \sigma_{\text{s}})$, as 
${\tilde {\mathbf x}}= {\mathbf x}/\mu$, we  obtain the
 Hamiltonian 
${\mathcal H} = 2 \xi \sum_{i=1}^{N} \ln ({\tilde r}_i/\tilde R)+
\Xi \sum_{\langle ij \rangle} 1/|{\tilde {\mathbf x}}_i-{\tilde {\mathbf x}}_j|$ (per $k_{\text{B}}T$),
where $\ell_{\text{B}} = e^2/(4 \pi \varepsilon \varepsilon_0 k_{\text{B}}T)$ is the 
Bjerrum length, and $\tilde{r}$ is the radial distance from the cylinder axis.
The  {\em coupling parameter},
$\Xi=2\pi q^3\ell_{\text{B}}^2\sigma_{\text{s}}$, 
is an indicator for the importance of ionic correlations: in the limit $\Xi \rightarrow 0$,
correlations are unimportant and MFT becomes exact; in the converse strong-coupling
limit $\Xi\rightarrow \infty$, MFT breaks down \cite{Andre}. 
The so-called  {\em Manning parameter} 
(rescaled inverse temperature), $\xi= q \ell_{\text{B}} \tau$, regulates the CCT and is a 
measure of counterion binding: 
According to MFT \cite{Manning,Alfrey,Zimm},  the CCT occurs 
for $\Delta=\ln (D/R)\rightarrow\infty$ at  the MF critical threshold
$\xi^{\text{MF}}_c=1$, above which a fraction $1-1/\xi$ of counterions condenses. 
To investigate the critical limit for large lateral system size, we introduce a 
(centrifugal) sampling method by 
mapping the radial coordinate to the logarithmic scale as $y=\ln (r / R)$.
The partition function  transforms as 
${\mathcal Z} \sim \int \left[\prod_{i}{\tilde r}_i{\text{d}}{\tilde r}_i {\text{d}}{\tilde z}_i{\text{d}}\phi_i \right] 
\exp(-{\mathcal H}) \sim \int \left[\prod_{i} {\text{d}} y_i {\text{d}}{\tilde z}_i {\text{d}}\phi_i \right]
 \exp(-{\mathcal H}_{\text{MC}} )$, where the  ``Hamiltonian"
\begin{equation}
 {\mathcal H}_{\text{MC}}=
 2 (\xi-1) \sum_{i=1}^{N} y_i+
 \Xi \sum _{\langle ij \rangle} 1/ |{\tilde {\mathbf x}}_i-{\tilde {\mathbf x}}_j|
\label{Hvirt}
\end{equation}
is used for Monte-Carlo sampling; it features a linear potential  (first term) acting on counterions from
competing energetic ($\sim 2\xi y$) and entropic or centrifugal ($\sim 2 y$) 
contributions associated with the cylindrical boundary.

In Fig.~\ref{fig:fig1}a, we show the snapshot-topviews  from our simulations for 
$\Delta =  \ln(D/R)=100$. De-condensation phase is reproduced for
small Manning parameter, $\xi=0.7$,  as counterions gather at the
outer boundary; while for $\xi=3$,  a fraction of  counterions accumulates or 
condenses around the  central cylinder. The transition regime for  intermediate $\xi$
exhibits strong finite-size effects: As seen for $\xi=1$ in Figs.~\ref{fig:fig1}a and b, 
only for large {\em logarithmic} system size, 
$\Delta\gg 1$, does de-condensation occur. This  is also demonstrated by vanishing  
radial distribution function of counterions, $p(\tilde r)$, for 
growing $\Delta$ in Fig. \ref{fig:fig1}c. For small coupling parameter $\Xi = 0.1$, 
the data for $p(\tilde r)$ compare well with the normalized MFT profile (solid curves)   
\begin{equation}
   p_{\text{MF}}({\tilde r})= \frac{\beta^2}{{2 \pi \xi \tilde r}^2}
              \sin^{-2}\left[ \beta \ln \frac{{\tilde r}}{{\tilde R}}
                               +\cot^{-1} \left( \frac{\xi-1}{\beta} \right) \right], 
\label{eq:rho_MF}
\end{equation}
where $\xi \geq \Delta /(1+\Delta)$ and
$\beta$ is given  by 
$\xi = ( 1+\beta^2)/(1-\beta\cot(-\beta\Delta))$ \cite{Alfrey}. 
Conversely, for large coupling $\Xi\rightarrow \infty$, 
 strong-coupling (SC) theory \cite{Andre} becomes valid and yields 
\begin{equation}
p_{\text{SC}}({\tilde r}) = \frac{2(\xi-1)}{2\pi \xi^2}
\bigg[1-e^{-2(\xi-1)\Delta}\bigg]^{-1}
  	 \left(\frac{\tilde r}{\tilde R}\right)^{-2\xi}.
\label{eq:rho_SC}		
\end{equation}
The break-down of MFT for increasing coupling parameter is demonstrated in
Fig.~\ref{fig:fig1}d (at fixed $\xi=3$), where the distribution 
function exhibits a gradual crossover 
between the asymptotic MFT and SC predictions \cite{Andre}. 
Enhanced ionic correlations at elevated $\Xi$ cause a remarkably larger
counterion density near the charged cylinder \cite{Deserno,Naji}. 
The central question is whether  these correlations influence 
the critical behavior associated with the CCT.

The standard analysis of critical behavior relies
on suitably-defined  order parameters \cite{critical}. 
Here, this role is taken by
 the inverse moments of the counterion distribution function, 
$S_n=\langle 1/{\tilde r}^n\rangle$ (with $n>0$); $S_n$ vanishes 
in the de-condensation phase,  but
attains a finite value in the condensation phase.
The MFT distribution  (\ref{eq:rho_MF})  gives 
\begin{equation}
   S_n^{\text{MF}}=\frac{\beta^2}{\xi^{n+1}}
             \int_0^\Delta {\text{d}}y \, e^{-ny} 
                \sin^{-2}\left[\beta y +\cot^{-1}\left(\frac{\xi-1}{\beta}\right)\right],
\label{eq:S_n_MF}
\end{equation} 
while in  the strong-coupling limit,  we have from Eq. (\ref{eq:rho_SC})
\begin{equation}
   S_n^{\text{SC}}=\frac{2(\xi-1)}{\xi^n(2\xi-2+n)}\frac{1-e^{-(2\xi-2+n)\Delta}}{1-e^{-2(\xi-1)\Delta}}.
 \label{eq:Sn_SC_full}
\end{equation}
The data for $S_1$ (the mean inverse localization length) in Fig.~\ref{fig:fig1}e
exhibit the condensation ($S_1>0$) and de-condensation ($S_1=0$) phases for a
wide range of couplings, $\Xi$. To locate the critical Manning parameter, $\xi_c$,
a finite-size analysis is required since criticality is masked both by 
 finite system size,  $\Delta$, and finite particle number, $N$, in the 
simulations (i.e. $S_1$ does not vanish and saturates
at a small value at critical point). 
%
%
\begin{figure*}
\includegraphics[width=17.5cm]{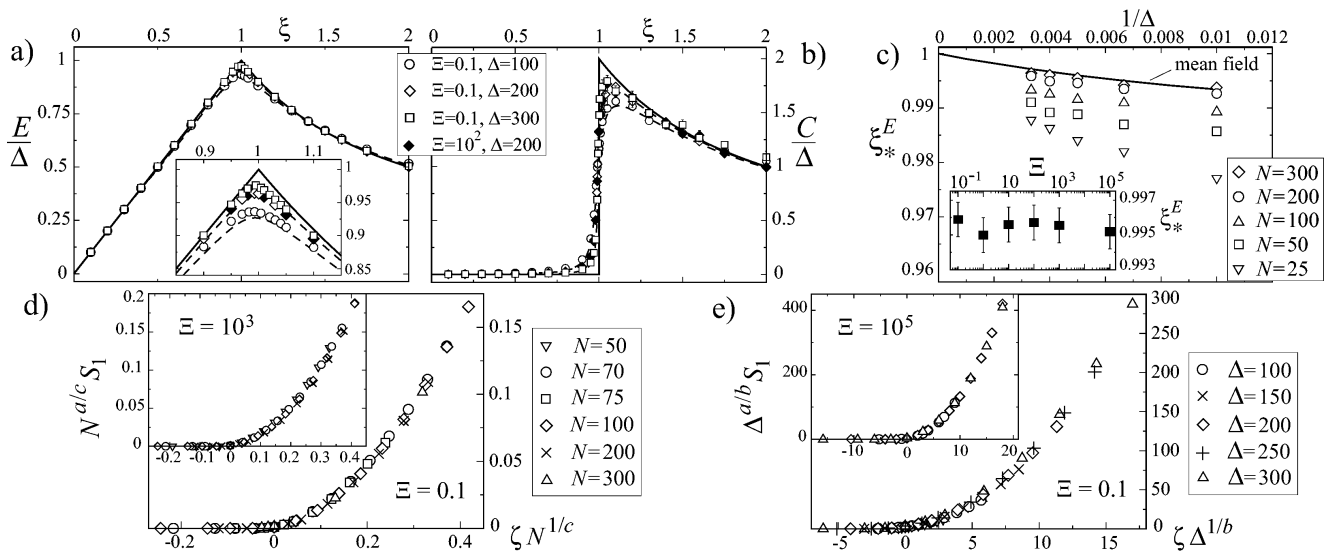}
\caption{a) Rescaled (potential)  energy, $E / \Delta$, 
and b) rescaled (excess) heat capacity,
$C/ \Delta $, as a function of Manning parameter, $\xi$, 
for fixed $N=100$ 
compared with  the  MFT prediction for $\Delta\rightarrow \infty$ (solid line), 
$\Delta=300$ and 100 (broken lines, from top to bottom) \cite{Naji}. 
c) Location of the peak of the energy, $\xi_\ast^E$, as a function of
$1/\Delta$ for  $\Xi=0.1$ compared with MFT  (solid line);
Inset: Peak location versus  $\Xi$ for $N=200$ and $\Delta=300$.
d) Scale invariance of the order parameter, $S_1$, near $\xi_c$ with respect to the reduced
Manning parameter, $\zeta  = 1-\xi_c/\xi$, and the counterion number, $N$,
with exponents  $a/c=2/3$ and $1/c=1/3$ (lateral system size $\Delta=300$). 
e) Scale invariance of $S_1$ for various $\Delta$  
with exponents $a/b=2$ and $1/b=1$ ($N=200$ is fixed).}
\label{fig:fig2}
\vspace{-0.4cm}
\end{figure*}
%
%
To this end, we study the singular behavior of rescaled energy 
$E=\langle {\mathcal H}/(Nk_{\text{B}}T)\rangle$ and heat capacity 
$C=\langle ({\delta \mathcal H}/k_{\text{B}}T)^2\rangle/N$, 
where ${\delta \mathcal H}={\mathcal H}-\langle {\mathcal H} \rangle$.
Simulation results for $E$ and $C$ (Figs.~\ref{fig:fig2}a, b) 
show a non-monotonic behavior 
that can be understood for large systems, $\Delta\gg 1$, using a simple
argument:  
for small $\xi$, all  counterions are unbound and the electrostatic
potential experienced by counterions 
is that of a bare cylinder, $\psi ({\tilde r}) = 2\xi\ln ({\tilde r}/{\tilde R})$ 
(per $k_{\text{B}}T/e q$), at  the outer boundary. 
The energy follows by a charging process as 
$ E  \simeq \psi ({\tilde D})  /2\simeq  \xi\Delta$. From the thermodynamic relation 
$\xi \partial  E / \partial \xi=  E- C $, one obtains $ C\simeq 0$.
For large $\xi$, 
the cylinder potential is screened by condensed counterions.
Using the MFT relation $p \sim e^{-\psi }$, 
 Eq. (\ref{eq:rho_MF}) gives
 $\psi ({\tilde r})  \simeq  2\ln {\tilde r}$. A fraction $1/\xi$
of counterions is de-condensed \cite{Manning,Zimm,Naji} and yields the leading
contribution to the energy; hence $ E\simeq  \psi (\tilde D) /(2 \xi)  \simeq  \Delta / \xi$ 
and $C\simeq 2\Delta/\xi$. Thus both quantities decay with 
$\xi$ as counterions
become increasingly condensed.
The energy exhibits a peak and the heat capacity develops a jump,
which become singular for an infinite system 
($\Delta\rightarrow \infty$, $N\rightarrow \infty$)
reflecting the CCT point, $\xi_c$  (Figs.~\ref{fig:fig2}a, b). 
We determine $\xi_c$ from the location of the energy peak, 
$\xi_\ast^E(\Delta, N)$, for increasing $N$ and $\Delta$.  
The numerical results for $\xi_\ast^E(\Delta, N)$ in Fig.~\ref{fig:fig2}c, 
obtained using the identity  $\xi \partial E/\partial \xi=E-C$,
compare favorably with the MF prediction (solid curve)  for small 
coupling $\Xi=0.1$ and increasing $N$.
The MF prediction asymptotically tends to the MF
threshold  $\xi^{\text{MF}}_c=1$ according to 
$\xi_c^{\text{MF}}-\xi_\ast^{E, {\text{MF}}}(\Delta) \sim 1/\Delta$
as $\Delta \rightarrow \infty$.
The location of the energy peak (and also the heat capacity jump) shows no 
dependence on the coupling parameter within error bars (Fig.~\ref{fig:fig2}c inset). 
We thus find a {\em universal} counterion-condensation threshold as 
$\xi_c=1.00\pm0.002$. 

We now turn to the near-threshold scaling behavior of the order parameter.
In the thermodynamic infinite-dilution limit 
($N\rightarrow\infty$, $\Delta\rightarrow \infty$),
 $S_n$ exhibits a power-law behavior as $S_n\sim \zeta^{\chi}$ 
 (close to and above $\xi_c$), where
$\zeta=1-\xi_c/\xi$ is the reduced Manning parameter (reduced temperature). 
Within MFT, we obtain $\chi_{\text{MF}}=2$ for all $n$  from Eq. (\ref{eq:S_n_MF}).
At criticality, $\zeta=0$,
one expects the scaling  $S_n\sim \Delta^{-\gamma}$  for 
$N\rightarrow \infty$ but finite $\Delta$ 
(within MFT, we obtain  $\gamma_{\text{MF}}=2$); 
while for $\Delta\rightarrow \infty$ but finite $N$, 
one expects  $S_n\sim N^{-\nu}$ ($\nu$ 
is not defined in MFT).  These relations indicate that 
close to criticality, $S_n(\zeta, \Delta, N)$ takes a 
{\em homogeneous scale-invariant} form, i.e. for $\lambda>0$, 
\begin{equation}
  S_n(\lambda\zeta, \lambda^{-b}\Delta, \lambda^{-c} N)=
        \lambda^{a}S_n(\zeta, \Delta, N)
\label{eq:GFSS1}
\end{equation}
where $a, b$, $c$ are related to the exponents $\gamma$, $\nu$,  $\chi$.
Choosing the scale factor as $\lambda=N^{1/c}$, one finds 
$  S_n(\zeta, \Delta, N)=N^{-a/c} S_n(\zeta N^{1/c}, \Delta N^{-b/c},1)$.
For large $\Delta N^{-b/c}$, as is the case in our simulations, 
and at the transition, $\zeta=0$, 
we have $S_n\sim N^{-a/c}$ and thus obtain $\nu=a/c$. 
For $N\rightarrow \infty$, a similar argument leads to
 $S_n\sim \zeta^a$,  which gives
$\chi=a$. 
In Fig.~\ref{fig:fig2}d we plot  the rescaled order parameter $ N^{a/c} S_1$ as a
function of $\zeta N^{1/c}$ for 
various $N$ and  fixed large $\Delta$.
By choosing the scaling exponents as $\nu =a/c = 2/3\pm 0.1$
and $\chi =a =2 \pm 0.4$,  we obtain excellent
data collapse both for small coupling $\Xi$
(main figure) and large $\Xi$  (inset). 
Choosing $\lambda=\Delta^{1/b}$ in Eq. (\ref{eq:GFSS1})
yields the  relation
$S_n(\zeta, \Delta, N)=\Delta^{-a/b} S_n(\zeta\Delta^{1/b}, 1, N\Delta^{-c/b})$. 
In Fig.~\ref{fig:fig2}e we plot $ \Delta^{a/b} S_1$ versus $\zeta\Delta^{1/b}$.
Good data collapse is obtained for $\gamma= a/b =2 \pm 0.6$
both at small (main figure) and large  $\Xi$ (inset),
which also demonstrates approximate independence from the scaling argument
$ N\Delta^{-c/b}$ \cite{Naji}. 
Our numerical results give the same critical exponents,
$\gamma$, $\nu$ and $\chi$,  for all $n$ and 
agree with MFT predictions (for $\gamma$ and $\chi$).
The exponents appear to be universal, i.e., independent 
of the coupling parameter, $\Xi$.

\begin{figure}
\includegraphics[width=7cm]{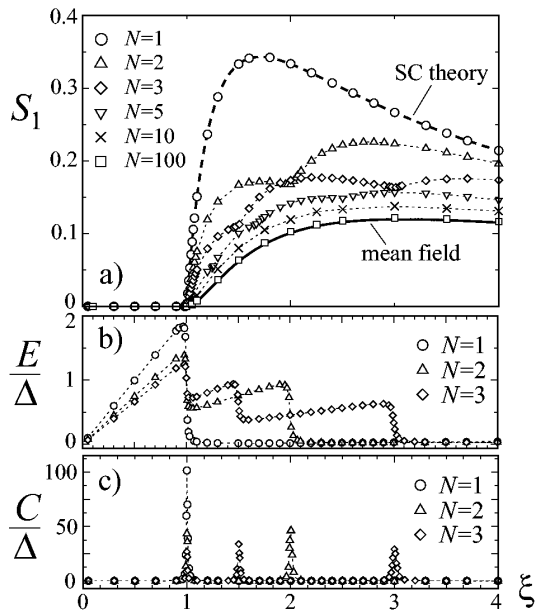}
\caption{ 2D results: 
a) Order parameter $S_1=\langle 1/{\tilde r}\rangle$ as a function 
of Manning parameter, $\xi$, for different particle numbers, $N$,  compared with
MFT  and SC  predictions, Eqs. (\ref{eq:S_n_MF}) and (\ref{eq:Sn_SC_full}).
b) Rescaled energy  and c) heat capacity (per particle) exhibit a set of singularities 
corresponding to successive condensation of single counterions 
for increasing $\xi$. Here $\Delta=300$. }
\label{fig:fig3}
\vspace{-0.4cm}
\end{figure}


Deviations from MFT in general grow with diminishing dimension \cite{critical}. 
The two-dimensional analogue of the counterion-cylinder system consists of 
logarithmically interacting mobile counterions and a central charged disk.
The 2D Hamiltonian reads
${\mathcal H}_{N} = 2\xi \sum_{i=1}^{N} \ln {\tilde r}_i -
2\Xi \sum_{\langle ij \rangle} \ln(|{\tilde {\mathbf x}}_i-{\tilde {\mathbf x}}_j|)$.
Unlike in 3D,  $\xi$ and $\Xi$ are related due to 
electroneutrality as $\Xi= \xi /N$. 
Thus the striking feature in 2D is that for a given 
Manning parameter, $\xi$, the coupling parameter tends to zero,
$\Xi\rightarrow 0$, in the limit  of many counterions,
 $N\rightarrow \infty$,  and MFT should become exact. 
Fig.~\ref{fig:fig3}a shows the simulated 
order parameter $S_1$ in 2D. 
For $N=1$, the data trivially follow the SC prediction (\ref{eq:Sn_SC_full}),   
dashed curve, and
for increasing $N$, they tend to the MF prediction (\ref{eq:S_n_MF}),  solid curve.
Accordingly,
scaling analysis of the condensation threshold and the
critical exponents  for $N \rightarrow \infty$ gives
identical results as in 3D and thus {\em no deviations
from MFT}. 
However, closer inspection of the 2D data in Fig.~\ref{fig:fig3}a 
reveals  a peculiar set of cusp-like singularities  for finite $N$.  In fact,
these singularities correspond to delocalization events of individual counterions, 
which give rise to a sawtooth-like structure for mean energy and 
a series of discrete peaks for heat capacity (Figs.~\ref{fig:fig3}b, c). 
This can be understood by a simple analysis of the 2D partition function:
Suppose  that  $N-m$ counterions  are firmly bound
to the central cylinder (disk), while $m-1$ counterions have 
evaporated to infinity   ($m=1,\ldots, N$). Neglecting the delocalized ions,
the partition function can be written as
${\mathcal Z}_N=\int \left[\prod_{i=m+1}^{N}  {\text{d}}^2 {\tilde x}_i\right] \exp(-{\mathcal H}_{N-m})
\times {\mathcal Z}_N^{(m)}$, where 
${\mathcal Z}_N^{(m)}=\int  {\text{d}}^2 {\tilde x}
\exp[-2\xi \ln {\tilde x}+\frac{2\xi}{N}\sum_{i=m+1}^{N}\ln 
(|{\tilde {\mathbf x}}_i-{\tilde {\mathbf x}}|)]$ 
is the contribution from the $m$-th counterion which is
assumed to be weakly localized.
It is thus de-correlated from the firmly 
bound ions and  ${\mathcal Z}_N^{(m)} $ approximately factorizes as
${\mathcal Z}_N^{(m)} \simeq  \int  {\text{d}}^2 {\tilde x} 
\exp[-2\xi \ln {\tilde x}+({2\xi}/{N})\sum_{i=m+1}^{N}\ln 
|{\tilde {\mathbf x}}|] \sim e^{(2-2 m \xi /N)\Delta}$. 
In the limit $\Delta\rightarrow \infty$,
${\mathcal Z}_N^{(m)}$ 
diverges for Manning parameters $\xi\leq N/m$.
 For $N$ counterions this gives a discrete set of singularities at $\xi_m = N/m$
 with diverging heat capacity $C\sim (\xi-\xi_m)^{-2}$, and 
 in agreement with our simulations (Fig.~\ref{fig:fig3}). 
For a  Manning parameter range
 $N/(m+1) <\xi< N/m$, there are $m$ de-condensed ions.
In the thermodynamic limit $N\rightarrow \infty$, the fraction
of de-condensed ions, $m/N$, becomes a smooth function and tends to 
the MF prediction  \cite{Manning,Zimm}, i.e., $m/N\rightarrow 1/\xi$. 

In summary, both in 2D and 3D  the location and critical 
exponents of the counterion-condensation transition at charged cylinders
are correctly described by mean-field theory. The heat capacity is
 experimentally  accessible: in 2D (parallel charged polymers), 
 it consists of a discrete set of peaks at which single counterions condense,
 and in the limit $N\rightarrow \infty$, it converges to the 3D shape
with a universal jump at the condensation threshold.

Discussions with H. Boroudjerdi, Y. Burak, H. Orland, Y. Y. Suzuki and support
from the DFG (SFB-486) are acknowledged.

\end{document}